\documentclass{ws-procs9x6-cpt16}

\begin{document}

\title{Extending the Graviton Propagator with a Lorentz-Violating Vector Field}

\author{Michael D. Seifert}

\address{Department of Physics, Astronomy, \& Geophysics, Connecticut
  College \\ 270 Mohegan Ave., New London, CT 06375 \\ 
  E-mail: mseifer1@conncoll.edu}

\begin{abstract}
  
  I discuss progress towards ``bootstrapping'' a Lorentz-violating
  gravity theory: namely, extending a linear Lorentz-violating theory
  of a rank-2 tensor to a non-linear theory by coupling this field to
  its own stress-energy tensor.
 
\end{abstract}

\bodymatter

The gravitational sector of the Standard Model Extension (SME) has
become of great interest in recent years, particularly with the recent
detection of gravitational waves by the LIGO Collaboration. The
treatment of gravity by the SME differs in an important way from its
treatment of other sectors. In the context of a gravitationally curved
spacetime, Lorentz violation cannot be thought of as due to a fixed
background tensor; instead, the Lorentz-violating tensor field
(represented abstractly by $\Psi^{\dots}$) must have its own
dynamics.\cite{AKnogo} The gravitational sector for the SME must
therefore be thought of as including the Einstein-Hilbert action,
various dynamical terms for $\Psi^{\dots}$, and small coupling terms between
$\Psi^{\dots}$ and the Riemann tensor.

The great majority of the work thus far in the gravitational sector of
the SME\cite{QBAKgrav,JTAKgravwave,MMAKgravwave} has focused on
linearized perturbations about a solution where spacetime is flat
($g_{ab} = \eta_{ab}$) and the Lorentz-violating tensor field is
constant ($\nabla \Psi^{\dots} = 0$).  In this limit, the dynamics of
$\Psi^{\dots}$ do not greatly affect the gravitational
dynamics.\cite{QBAKgrav} However, in strongly curved spacetimes, a
constant tensor field $\Psi^{\dots}$ will in general not exist.  If
the SME framework is to address such spacetimes (such as compact
objects, black holes, or cosmological spacetimes), we will have to
address the dynamics of the underlying tensor field $\Psi^{\dots}$.


An old idea in the context of gravitational physics is the idea of
``bootstrapping'' gravity from a linear theory to a non-linear
theory.\cite{OGbootstrap} This method is based on the idea that
``gravity gravitates'': the stress-energy \emph{of} the gravitational
field should act as a source \emph{for} the gravitational field.  This
idea predates the SME framework by a few decades, and it is
instructive to ask whether one can extend this idea to situations with
violation of Lorentz symmetry.  In this process, three main questions
arise:
\begin{enumerate}

\item At the level of a linear free-field gravity theory, what kinds
  of theories can I write down if I allow for violations of Lorentz
  symmetry?  

\item Can such linear theories be bootstrapped to non-linear theories?

\item Does requiring that the linear theory be ``bootstrappable''
  place constraints on the dynamics of the Lorentz-violating field
  $\Psi^{\dots}$?

\end{enumerate}

Let us consider a general linear theory of a source-free symmetric
rank-two tensor field $h_{ab}$ in flat spacetime, with action
\begin{equation}
S = \frac{1}{2} \int d^4 x \, \partial_a h_{bc} \mathcal{P}^{abcdef} \partial_d
h_{ef} \label{actioneq}
\end{equation}
where the \emph{propagator tensor} $\mathcal{P}^{abcdef}$ is some
constant tensor to be determined.  The resulting equations of motion
are then
\begin{equation}
  \left( \mathcal{P}^{abcdef} +
    \mathcal{P}^{aefdbc} \right) \partial_a \partial_d h_{ef} = 0. \label{EOMs}
\end{equation}
By symmetry in \eqref{actioneq}, we can assume that
$\mathcal{P}^{abcdef}$ is symmetric under the exchanges
$b \leftrightarrow c$, $e \leftrightarrow f$, and
$\{abc\} \leftrightarrow \{def\}$.  From Eq.\ \eqref{EOMs}, we can
also take the propagator to be symmetric under the exchange
$\{bc\} \leftrightarrow \{ef\}$.  Finally, we will want to insert a
conserved stress-energy tensor as a source on the right-hand side of
\eqref{EOMs}; this implies that the divergence of the left-hand side
of \eqref{EOMs} must also vanish. In Fourier space, this implies that
the quantity $\mathcal{P}^{abcdef} k_a k_b k_d = 0$ for all choices of
wave propagation vector $k_a$.

We expect that in the end, $\mathcal{P}^{abcdef}$ will not be a
fundamental object but rather a function of simpler tensors, such as a
``fiducial'' flat metric $\eta^{ab}$ or a Lorentz-violating tensor
field of some kind.  The strategy is then to write down the most
general $\mathcal{P}^{abcdef}$ that can be constructed out of these
simpler tensors, subject to the above symmetry constraints.  Using the
fiducial metric alone, for example, we find that the unique propagator
satisfying the desired symmetry properties is the usual
Lorentz-invariant linearized gravity propagator, as expected:
\begin{eqnarray}
  (\mathcal{P}_{LI})^{abcdef} 
  &=& \eta^{a(b} \eta^{c)d} \eta^{ef} + \eta^{a(e} \eta^{f)d}
      \eta^{bc} - \eta^{a(b} \eta^{c)(e} 
      \eta^{f)d} \nonumber \\
  && \qquad  {}- \eta^{a(e} \eta^{f)(b}
     \eta^{c)d} - \eta^{ad} \eta^{bc}
     \eta^{ef} + \eta^{ad} \eta^{b(e} \eta^{f)c}. 
\end{eqnarray}

The simplest possible Lorentz-violating tensor field $\Psi^{\dots}$
would be a four-vector field $A^a$. If we follow the above procedure,
constructing the propagator out of $\eta^{ab}$ and $A^a$, we find that
the resulting Lorentz-violating propagator
$(\mathcal{P}_{LV})^{abcdef}$ only has one free parameter $\xi$.
What's more, this $(\mathcal{P}_{LV})^{abcdef}$ is equivalent to the
Lorentz-invariant $(\mathcal{P}_{LI})^{abcdef}$ under the substitution
\begin{equation}
  \eta^{ab} \to \tilde{\eta}^{ab} \equiv \eta^{ab} + \xi A^a A^b.
\end{equation}
In other words, the introduction of a Lorentz-violating vector field
only allows one to change the ``effective metric'' $\tilde{\eta}^{ab}$
for linearized gravity. This will have the effect of changing the
``light cones'' for gravitational wave propagation; however, it will
not allow for more exotic effects such as dispersion or birefringence
of gravitational waves.  This result is in agreement with the more
general results of Ref.\ \refcite{MMAKgravwave}.

Having classified the ways in which the linearized theory can break
Lorentz symmetry, I now turn to the question of extending it to a
non-linear theory.  To do this, I follow the method of
Deser,\cite{Deserbootstrap, RPAKbootstrap} and write down a
\emph{first-order} linear model in terms of a \emph{densitized} tensor
$\mathfrak{h}^{ab}$ and an auxiliary tensor $\Gamma^a {}_{bc}$.  In
the Lorentz-invariant case, the action for this model is
\begin{equation}
  S = \int d^4 x \, \left[ 2 \mathfrak{h}^{ab} \partial_{[c} \Gamma^c
    {}_{b]a} + 2 \eta^{ab} \Gamma^c
    {}_{d [c} \Gamma^d {}_{a]b} + \mathcal{L}_\text{mat} (\eta^{ab},
    A^a, \partial_a A^b) \right]. \label{bootstrapLI}
\end{equation}
The equations of motion for $\mathfrak{h}^{ab}$ and
$\Gamma^a {}_{bc}$, combined, are equivalent to the linearized
vacuum Einstein equations if we interpret $\mathfrak{h}^{ab}$ as the
perturbation to the densitized inverse metric:
$\mathfrak{g}^{ab} = \eta^{ab} + \mathfrak{h}^{ab}$.  Under this
interpretation, $\mathfrak{h}^{ab}$ must be coupled to the
trace-reversed stress-energy
$\tau_{ab} = \delta \mathcal{L}/\delta \eta^{ab}$.  The second term in
\eqref{bootstrapLI}, along with the matter Lagrangian
$\mathcal{L}_\text{mat}$, contribute to the stress-energy (note that
the first term is independent of $\eta^{ab}$.)  We will thus need to
add two new coupling terms to the action \eqref{bootstrapLI}:
\begin{equation}
  S \to S + \int d^4 x \, \mathfrak{h}^{ab} \left[ 2 \Gamma^c
    {}_{d [c} \Gamma^d {}_{a]b} + (\tau_\text{mat})_{ab}
  \right]. \label{couplingLI} 
\end{equation}
Importantly, the new term in the gravitational sector (the first term
in \eqref{couplingLI}) does not depend on $\eta^{ab}$, and so the
bootstrap procedure terminates here for the gravitational sector.  In
the matter sector, the term $(\tau_\text{mat})_{ab}$ may itself depend
on the metric $\eta^{ab}$, and so the contributions to the
stress-energy from this coupling term must be added in as well.  The
iteration of this procedure can, in principle, generate an infinite
series of terms.  However, assuming that various integrability
conditions are satisfied,\cite{RPAKbootstrap} the resulting series can
be summed up to yield a matter action that is minimally coupled to the
densitized metric $\mathfrak{g}^{ab}$.  The gravitational terms,
meanwhile, combine into the Palatini action for general relativity:
\begin{equation}
  S = \int d^4 x \, \left[ \mathfrak{g}^{ab} R_{ab} [ \Gamma ] +
    \mathcal{L}_\text{mat} (\mathfrak{g}^{ab}, 
    A^a, \nabla_a A^b) \right], \label{bootstrappedLI}
\end{equation}
where the Ricci tensor $R_{ab}$ is viewed here as a function of the
connection coefficients $\Gamma$.

Perhaps surprisingly, this scenario changes very little when we relax
the assumption of Lorentz symmetry.  As found above, the only possible
modification that can be made to the linearized gravity action in the
presence of a Lorentz-violating vector field is to
replace the matter metric $\eta^{ab}$ with an effective metric
$\tilde{\eta}^{ab} = \eta^{ab} + \xi A^a A^b$.  This will give rise to
a new term $\xi A^a A^b \Gamma^c {}_{d [c} \Gamma^d {}_{a]b}$ in the
action; but this term is independent of $\eta^{ab}$, and so does not
contribute to the stress-energy tensor.  Thus, the entire bootstrap
procedure carries through as before; the only difference is that the
densitized metric that appears in the Palatini action is not the same
as that appearing in the matter action:
\begin{equation}
  S = \int d^4 x \, \left[ \tilde{\mathfrak{g}}^{ab} R_{ab} [ \Gamma ] +
    \mathcal{L}_\text{mat} (\mathfrak{g}^{ab}, 
    A^a, \nabla_a A^b) \right] \label{bootstrappedLV}
\end{equation}
where $\tilde{\mathfrak{g}}^{ab} \equiv \mathfrak{g}^{ab} + \xi A^a
A^b$.

This action could then be rewritten using the (undensitized)
gravitational metric as a fundamental variable; the result would be
some kind of exotic tensor-vector theory of gravity.  It is important
to note, however, that the construction of this theory required that
the action for the Lorentz-violating field $A^a$ itself be amenable
to ``bootstrapping''; in particular, it must satisfy various
integrability constraints at each stage of the bootstrap procedure.  I
conjecture that a symmetry-breaking potential and a ``Maxwell-type''
kinetic term for $A^a$ will satisfy these constraints, and that more
exotic kinetic terms will fail; but this has not yet been proven.

\section*{Acknowledgments}
I would like to thank Connecticut College for their financial support in
attending this conference.

\end{document}